%% file: main.tex
\documentclass[conference]{IEEEtran}
\IEEEoverridecommandlockouts
\usepackage{cite}
\usepackage{amsmath,amssymb,amsfonts}
\usepackage{algorithmic}
\usepackage{graphicx}
\usepackage{textcomp}
\usepackage{xcolor}
\def\BibTeX{{\rm B\kern-.05em{\sc i\kern-.025em b}\kern-.08em
    T\kern-.1667em\lower.7ex\hbox{E}\kern-.125emX}}
    
\usepackage[left=2.54cm,right=2.54cm,top=2.54cm,bottom=2.54cm]{geometry}

\begin{document}


\title{Enabling Reproducible Analysis of Complex Workflows on the Edge-to-Cloud Continuum}

\author{\IEEEauthorblockN{
\textbf{PhD Student:} Daniel Rosendo\IEEEauthorrefmark{1}
\textbf{Advisors:} Alexandru Costan\IEEEauthorrefmark{1},
Gabriel Antoniu\IEEEauthorrefmark{1},
Patrick Valduriez\IEEEauthorrefmark{2}}
\IEEEauthorblockA{\IEEEauthorrefmark{1}University of Rennes, Inria, CNRS, IRISA - Rennes, France}
\IEEEauthorblockA{\IEEEauthorrefmark{2}University of Montpellier, Inria, CNRS, LIRMM - Montpellier, France\\
\{daniel.rosendo, alexandru.costan, gabriel.antoniu, patrick.valduriez\}@inria.fr}\\
}

\maketitle


\begin{abstract}

Distributed digital infrastructures for computation and analytics are now evolving towards an interconnected ecosystem allowing complex applications to be executed from IoT Edge devices to the HPC Cloud (aka the \emph{Computing Continuum}, the \emph{Digital Continuum}, or the \emph{Transcontinuum}). Understanding end-to-end performance in such a complex continuum is challenging. This breaks down to reconciling many, typically contradicting application requirements and constraints with low-level infrastructure design choices. One important  challenge is to accurately reproduce relevant behaviors of a given application workflow and representative settings of the physical infrastructure underlying this complex continuum. We introduce a rigorous methodology for such a process and validate it through \textbf{E2C}\textit{lab}. It is the first platform to support the complete experimental cycle across the Computing Continuum: deployment, analysis, optimization. Preliminary results with real-life use cases show that \textbf{E2C}\textit{lab} allows one to understand and improve performance, by correlating it to the parameter settings, the resource usage and the specifics of the underlying infrastructure.

\end{abstract}


\input{1_introduction}

\input{2_e2clab}

\input{3_next_steps}


\bibliographystyle{IEEEtran}
\bibliography{references}

\end{document}

%% file: 1_introduction.tex
\section{Context}

The explosion of data generated from the Internet of Things (IoT) and the need for real-time analytics has resulted in a shift of the data processing paradigms 
towards decentralized and multi-tier computing infrastructures and services. New challenging application scenarios are emerging from a variety of domains such as healthcare, self-driving vehicles
, precision agriculture, \emph{etc.} 
This contributes to the emergence of what is called the \emph{Computing Continuum}~\cite{etp4-hpc-20}. It seamlessly combines resources and services at the center (e.g., in Cloud datacenters), at the Edge, and in-transit, along the data path. Typically data is first generated and preprocessed (e.g., filtering, basic inference) on Edge devices, while Fog nodes further process partially aggregated data. Then, if required, data is transferred to HPC-enabled Clouds for Big Data analytics, AI model training, and global simulations.

\section{Problem Statement}
Despite an always increasing number of dedicated systems for data processing on each component of the continuum,
this vision of ubiquitous computing remains largely unrealized. This is due to the complexity of deploying large-scale, real-life applications on such heterogeneous infrastructures, which breaks down to configuring a myriad of system-specific parameters and reconciling many requirements or constraints, e.g., in terms of communication latency, energy consumption, resource usage, data privacy. A first step towards reducing this  complexity and enabling the Computing Continuum vision is to enable a \textbf{holistic understanding of performance} in such environments. That is, finding a rigurous approach to answering questions like: \emph{(1) Which system parameters and infrastructure configurations impact on performance and how? (2) Where should the workflow components be executed to minimize communication costs and end-to-end latency?} 


\section{State of the art}

Approaches based on workflow modelling \cite{sadiq2004data} and simulation \cite{svorobej2019simulating} raise some important challenges in terms of specification, modelling, and validation in the context of the Computing Continuum. For example, it is increasingly difficult to assess the impact of the inherent complexity of hybrid Edge-Cloud deployments on performance. 
At this stage, experimental evaluation remains the main approach to gain accurate insights of performance metrics and to build precise approximations of the expected behavior of large-scale applications on the Computing Continuum, as a first step prior to modelling.

\section{Challenges}

A key challenge in this context is to be able to \textbf{reproduce in a representative way the application behavior in a controlled environment}, for extensive experiments in a large-enough spectrum of potential configurations of the underlying Edge-Fog-Cloud infrastructure.
In particular, this means rigorously mapping the scenario characteristics to the \textit{experimental environment}, 
identifying and controlling the relevant \textit{configuration parameters} of applications and system components, defining the relevant \textit{performance metrics}. 
The above process is non-trivial due to the multiple combination possibilities of heterogeneous hardware/software resources, system components for data processing, analytics or model training.

\section{PhD Objectives}

In order to allow other researchers to leverage the experimental results and advance knowledge in different domains, experimental methodologies need to enable three R's of research quality: \textbf{Repeatability, Replicability,} and \textbf{Reproducibility (3R's)}. This translates to establishing a \textit{well-defined experimentation methodology} and providing \textit{transparent access to the experiment artifacts} and \textit{experiment results}.

The Computing Continuum vision calls for a rigorous and systematic methodology to map real-world application components and dependencies to infrastructure resources,
a complex process that can be error prone. Key research goals are: \textit{1)} to identify relevant characteristics of the application workloads and of the underlying infrastructure as a means to enable accurate experimentation and benchmarking in relevant infrastructure settings in order to understand their performance; and \textit{2)} to ensure research quality aspects such as the 3R's.


%% file: 2_e2clab.tex
\section{Our Contribution: \textbf{E2C}\textit{lab}}
\textbf{E2C}\textit{lab}~\cite{rosendo2020e2clab} implements a methodology that supports the complete experimental cycle across the edge-to-cloud continuum, including deployment, configuration, optimization, and experiment execution in a reproducible way. It may be used by researches to deploy real-life applications on large-scale testbeds and perform meaningful experiments in a systematic manner. The \textbf{main contributions} of this work are:



A \textbf{rigorous methodology for designing experiments with real-world workloads on the Computing Continuum} spanning from the Edge to the Cloud; this methodology provides guidelines to move from real-world use cases to the design of relevant testbed setups for experiments enabling researchers to understand performance and to ensure the 3R's properties.
    
A novel \textbf{framework named} \textbf{E2C}\textit{lab} that implements this methodology and allows researchers to deploy their use cases on real-world large-scale testbeds, e.g., G5k~\cite{bolze:hal-00684943}. To the best of our knowledge, \textbf{E2C}\textit{lab} is the first platform to support the complete analysis cycle of an application on the Computing Continuum: \emph{(i)} the configuration of the experimental environment; \emph{(ii)} the mapping between the application parts and machines on the Edge, Fog and Cloud; \emph{(iii)} the deployment and monitoring of the application on the infrastructure; and \emph{(iv)} the automated execution and gathering of results.


    
A \textbf{large scale experimental validation} on the G5K~\cite{bolze:hal-00684943} testbed with Pl@ntNet~\cite{joly2016look}, a real-life use case. \textbf{E2C}\textit{lab} allows optimizing the Pl@ntNet's performance based on the analysis of the parameter settings and correlation to processing time and resource usage~\cite{rosendo2021reproducible}.

\section{Preliminary Results}
We illustrate~\cite{rosendo2020e2clab} \textbf{E2C}\textit{lab} usage with a \textbf{real-life Smart Surveillance System} deployed on the Grid'5000 testbed, showing that our framework allows one to understand how the Cloud-centric and the hybrid Edge-Cloud processing approaches impact performance metrics such as latency and throughput. Besides, we validate~\cite{rosendo2021reproducible} \textbf{E2C}\textit{lab} with \textbf{Pl@ntNet}, another \textbf{real-life use case}. We demonstrate that \textbf{E2C}\textit{lab} guides on the optimization of the Pl@ntNet performance based on the analysis of the parameter settings and correlation to processing time and resource usage. Preliminary results show that Pl@ntNet's deployment configurations found by \textbf{E2C}\textit{lab} perform better than the current ones used in the production servers.

%% file: 3_next_steps.tex
\section{Next Research Steps}
We are exploring \textbf{parallel and scalable optimization} techniques that supports surrogate modeling optimization for large-scale multi-objective optimization problems. In this direction, we have an ongoing collaboration with Argonne National Laboratory members, where we are discussing potential solutions to support the optimization of complex application workflows on the Edge-to-Cloud Continuum. Furthermore, since \textbf{E2C}\textit{lab} supports \textbf{reproducible experiments}, we will explore and propose techniques for \textbf{runtime provenance} collection in large-scale and distributed experimental environments. The goal is to provide additional context that more accurately explains the experiment execution and results. This research direction is a collaboration with the Federal University of Rio de Janeiro, Brazil.